\documentclass[12pt]{article}
\usepackage{graphicx}
\usepackage{amssymb}
\usepackage{graphics}
\usepackage{amsmath}
\usepackage{amsfonts}
\usepackage{bm}

\newcommand\pubnumber{SNSN-323-63}
\newcommand\pubdate{\today}

\def\napoli{Dipartimento di Fisica E.R. Caianiello, Universita' di Salerno, ITALY \\
INFN - Gruppo Collegato di Salerno, ITALY}
\def\support{\footnote{Work supported by INFN.}}

\def\Title#1{\begin{center} {\Large #1 } \end{center}}
\def\Author#1{\begin{center}{ \sc #1} \end{center}}
\def\Address#1{\begin{center}{ \it #1} \end{center}}

\newcommand\pubblock{\rightline{\begin{tabular}{l} \pubnumber\\
         \pubdate  \end{tabular}}}
\newenvironment{Abstract}{\begin{quotation}  }{\end{quotation}}
\newenvironment{Presented}{\begin{quotation} \begin{center}
             PRESENTED AT\end{center}\bigskip
      \begin{center}\begin{large}}{\end{large}\end{center} \end{quotation}}

\def\beq{\begin{equation}}
\def\eeq#1{\label{#1}\end{equation}}
\def\eeqn{\end{equation}}

\def\beqa{\begin{eqnarray}}
\def\eeqa#1{\label{#1}\end{eqnarray}}
\def\eeqan{\end{eqnarray}}

\let\bar=\overbar

\def\Dslash{\not{\hbox{\kern-4pt $D$}}}
\def\dslash{\not{\hbox{\kern-2pt $\del$}}}

\def\msb{{\bar{\ssstyle M \kern -1pt S}}}

\begin{document}
\begin{titlepage}
\pubblock

\vfill
\Title{Consequences of Modified Cosmologies in DM abundance and PeV IceCube signals}
\vfill
\Author{G. Lambiase\support}
\Address{\napoli}
\vfill
\begin{Abstract}
To explain the high-energy astrophysical neutrino flux with energies $\sim PeV$ recently observed by IceCube collaboration and the relic abundance of DM in the Universe, we consider the minimal 4-dimensional operator $\sim y_{\alpha \chi}\overline{{L_{L_{\alpha}}}}\, H\, \chi$. The cosmological background is assumed to evolve according to $f({\cal T})$ cosmology, where ${\cal T}$ is the scalar torsion.
\end{Abstract}
\vfill
\begin{Presented}
NuPhys2018, Prospects in Neutrino Physics\\
Cavendish Conference Centre, London, UK, December 19--21, 2018
\end{Presented}
\vfill
\end{titlepage}
\def\thefootnote{\fnsymbol{footnote}}
\setcounter{footnote}{0}

\section{Introduction}

The IceCube Collaboration \cite{[2]} has recently reported several events of neutrinos with energies of the order of PeV.
Candidates for the generation of such neutrino high energy events could be various astrophysical sources \cite{Sahu:2014fua}. Till now, however, there are no clear correlations with the known astrophysical sources (SNe remnants or AGN \cite{Aartsen:2016oji}). Other interesting ideas assume that neutrinos could arise from the decay of PeV mass Dark Matter (DM) \cite{merle}. Besides this topic, another interesting question is whether it is possible to explain both the PeV DM relic density (i.e. $\Omega_{DM}h^2\Big|_{obs}=0.1188\pm 0.0010$ \cite{Planck}) and the decay rate required for IceCube with only one operator. The minimal DM-neutrino 4-dimensional interaction we consider is $ y\, \bar L\cdot H \chi$ \cite{merle}. In the framework of the comsological standard model, this operator fails to account for both the PeV dark matter relic abundance and the decay rate required to explain IceCube. We therefore consider the possibility that Universe evolves according to modified cosmology that consist in an extension of Einstein's theory (see for example
\cite{DeFelice:2010aj}).
In these models, the expansion rates $H$ of the Universe can be written in terms of the expansion rate $H_{GR}$ of General Relativity (GR) \cite{fornengo}
 \begin{equation}\label{H=AHGRIce}
   H_{MC}(T)=A(T)H_{GR}(T)\,, \quad A(T)=\eta\left(\frac{T}{T_*}\right)^\nu\,,
 \end{equation}
where $T_*$ is a reference temperature, and $\{\eta, \nu\}$ free parameters that depend on the cosmological model under consideration \cite{fornengo}.
To preserve the successful predictions of BBN, one refers to the pre-BBN epoch since it is not directly constrained by cosmological observations.


\section{PeV neutrinos and modified cosmology}

The simplest 4-dimensional operator that allows to explain the IceCube high energy signal, is
 \begin{equation}\label{4dop}
 {\cal L}_{d=4}=y_{\alpha \chi}\overline{{L_{L_{\alpha}}}}\, H\, \chi\,, \qquad \alpha = e, \mu, \tau\,,
 \end{equation}
where $\chi$ is the DM particle that transforms as $\chi\sim (1,1,0)$ of SM, $H\sim (1, 2, +1/2)$ is the Higgs doublet, $L_{L_{\alpha}}\sim (1, 2, -1/2)$ is the left-handed lepton doublet corresponding to the generation $\alpha (=e, \mu\, \tau)$, and finally $y_{\alpha \chi}$ are the Yukawa couplings.
We consider the freeze-in production \cite{hall,merle}, i.e. the DM particles are never in thermal equilibrium since they interact very weakly, but are gradually produced from the hot thermal bath. As a consequence, a sizable DM abundance is allowed until the temperature falls down to $T\sim m_\chi$ (temperatures below $m_\chi$ are such that DM particles phase-space is kinematically difficult to access).

\begin{figure}[btp]
  \centering
  \includegraphics[width=8.0cm]{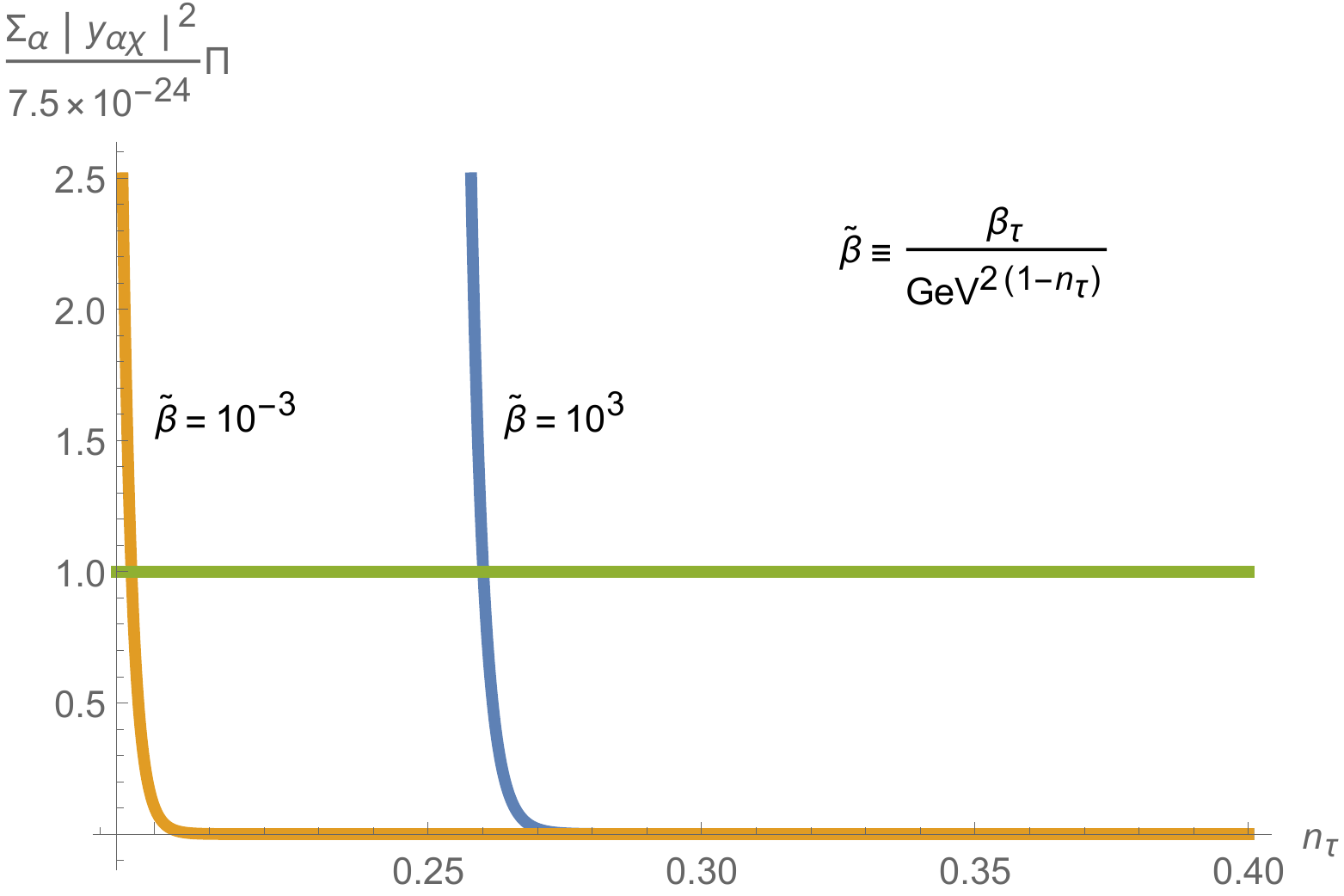}\\
  \caption{For fixed values of $\eta=1$ and $\beta_{\cal T}$, the values of $\Pi$ needed to explain DM relic abundance and IceCube data follow for $n_{\cal T}\lesssim 0.25$.}\label{PiTorsion}
\end{figure}

Denoting with $Y_\chi=n_\chi/s$ the DM abundance, where $n_\chi$ is the number density of the DM particles and $s=\frac{2\pi^2}{45}g_*(T)T^3$ the entropy density, the evolution of the DM particle is governed by the Boltzmann equation $\frac{dY_\chi}{dT}=-\frac{1}{HTs}\left[\frac{g_\chi}{(2\pi)^3}\int C\frac{d^3p_\chi}{E_\chi}\right]$
where $H$ is the expansion rate of the Universe and $C$ the general collision term. In modified cosmology,
the DM relic abundance is given by \cite{mohantyEPJC}
 \begin{equation}\label{DMeta}
   \Omega_{DM}h^2
  \simeq  0.1188 \frac{\sum_\alpha |y_{\alpha\chi}|^2}{7.5\times 10^{-24}}\, \Pi\,, \,\,\, \text{where} \,\,\,
    \Pi\equiv \frac{2^{3+\nu}}{3\pi \eta}\left(\frac{T_*}{m_\chi}\right)^\nu
   \Gamma\left(\frac{5+\nu}{2}\right)\Gamma\left(\frac{3+\nu}{2}\right)\,. \nonumber
  \end{equation}
with $\nu>-3$. $\Pi$ accounts for all corrections induced by modified cosmology.
To explain the DM relic abundance and the IceCube data, we require
 \begin{equation}\label{Phivalue}
   \frac{\sum_\alpha |y_{\alpha\chi}|^2}{7.5\times 10^{-24}}\, \Pi \lesssim {\cal O}(1)\,.
 \end{equation}
%
Consider the model $ S_I = \frac{1}{16\pi G}\int{d^4xe\left[{\cal T}+f({\cal T})\right]}$, where $f({\cal T})=\beta_{\cal T} |{\cal T}|^{n_{\cal T}}$ ($f({\cal T})$ is a generic function of the torsion ${\cal T}$). In such a case \cite{mohantyEPJC}, one gets (see Eq. (\ref{H=AHGRIce})) 
 \begin{equation}\label{ATTorsion}
   T_*\equiv \left(\frac{24\pi^3 g_*}{45}\right)^{\frac{1}{4}}
   (2n_{\cal T}+1)^{\frac{1}{4(1-n_{\cal T})}}\left(\frac{\beta_{\cal T}}{\mbox{GeV}^{2(1-n_{\cal T})}}\right)^{\frac{1}{4(1-n_{\cal T})}}\left(\frac{M_{Pl}}{\mbox{GeV}}\right)^\frac{1}{2}\mbox{GeV}\,,
 \end{equation}
$\eta=1$ and $\nu=\frac{2}{n_{\cal T}}-2$. For $a(t)=a_0 t^\delta$, i.e. $H=\frac{\delta}{t}$, it follows  $T(t)a(t)=constant$.
The transition temperature $T_*$ given in (\ref{ATTorsion}) has to be used into Eq. (\ref{Phivalue}).
In Fig. \ref{PiTorsion} is plotted (\ref{Phivalue}) for the $f({\cal T})$ model.
The value of $\beta_{\cal T}$ is obtained by fixing the transition temperature at $T_*\sim 10^{11}-10^9$GeV, that is $T_*\gg m_\chi$.

\section{Conclusions}

In this paper we have studied the possibility to reconcile the current bound on DM relic abundance with IceCube data in terms of the 4-dimensional operator (\ref{4dop}). We have shown that modified gravity models can explain the IceCube outputs and at the same time the DM relic abundance observed in a minimal particle physics model. We have considered cosmological models related to torsion ${\cal T}$.



\begin{thebibliography}{99}





\bibitem{[2]} M.G. Aartsen {\it et al}, {\it Phys. Rev. Lett} {\bf 115} 081102 (2015).
I. Cholis and D. Hooper, {\it JCAP} {\bf 06} 030 (2013).
L.A. Anchordoqui {\it et al}, {\it JHEAp} {\bf 1-2} 1 (2014).
\bibitem{Sahu:2014fua}  S. Sahu and L.S. Miranda, {\it Eur. Phys. J.} C {\bf 75} 273 (2015).
\bibitem{Aartsen:2016oji} M.G. Aartsen {\it et al}, {\it Astrophys. J.}  {\bf 835} 151 (2017).
\bibitem{merle} M. Chianese and A. Merle, {\it JCAP} {\bf 04} 017 (2017).
\bibitem{Planck} P.A.R. Ade {\it et al}, {\it Astron. Astrophys.} {\bf 594} 13 (2016).
\bibitem{hall} L.J. Hall, K. Jedamzik, J. March-Russell, and S.M. West, {\it JHEP} {\bf 03} 080 (2010).
\bibitem{DeFelice:2010aj} A. De Felice and S. Tsujikawa, {\it Living Rev. Rel.} {\bf 13} 3 (2010).
	G. Chakravarty, S. Mohanty, and G. Lambiase, {\it Int. J. Mod. Phys.} D {\bf 26} 1730023 (2017 ).
     R. Durrer and R. Maartens,  {\it Gen. Relat. Grav.} {\bf 40} 301 (2008).
\bibitem{fornengo} R. Catena, N. Fornengo, M. Pato, L. Pieri and A. Masiero, {\it Phys. Rev.} D {\bf 81} 123522 (2010).
    M. Kamionkowski and M.S. Turner,  {\it Phys. Rev.} D {\bf 42} 3310 (1990).
    P. Salati,  {\it Phys. Lett.} B {\bf 571} 121 (2003).
    G. Lambiase,  {\it JCAP} {\bf 10} 028 (2012).
     G. Gelmini and P. Gondolo,  arXiv:1009.3690. 
                G. D'Amico, M. Kamionkowski, and K. Sigurdson, arXiv:0907.1912. 
     L. Randal and R. Sundrum,  {\it Phys. Rev. Lett.} {\bf 83} 4690 (1991).
     F. Profumo and P. Ullio,  {\it JCAP} {\bf 0311} 006 (2003).
                F. Rosati,  {\it Phys. Lett.} B {\bf 570} 5 (2003).
\bibitem{mohantyEPJC} G. Lambiase, S. Mohanty, and A. Stabile,  {\it Eur. Phys. J.} C {\bf 78} 350 (2018).
\bibitem{fTEPJC} R. Ferraro and F. Fiorini,  {\it Phys. Rev.} D {\bf 75} 084031 (2007).
   R. Aldrovandi and J.G. Pereira,  {\it Teleparallel Gravity: An Introduction}, 2013 (Springer, Dordrecht).
   S. Capozziello, G. Lambiase and E.N. Saridakis,  {\it Eur. Phys. J.} C {\bf 77} 576 (2017).
\end{thebibliography}
\end{document}